\documentclass[aps,prl,reprint,twocolumn,superscriptaddress,showkeys,a4paper]{revtex4}
\usepackage{amsfonts}
\usepackage{amssymb}
\usepackage{bm, bbm, dsfont}
\usepackage[dvips]{graphicx}
\usepackage[T1]{fontenc}
\usepackage[latin2]{inputenc}
\usepackage{amsmath}
\usepackage{latexsym}

\usepackage{color}

\def\dt{{\rm d}\,}

\newcommand{\ket}[1]{| #1 \rangle}
\newcommand{\bra}[1]{\langle #1 |}

\def\duzomniejsze{<\kern-.7mm<}
\def\duzowieksze{>\kern-.7mm>}

\def\textbf#1{{\bf #1}}
\def\be{\begin{equation}}
\def\ee{\end{equation}}
\def\ben{\begin{eqnarray}}
\def\een{\end{eqnarray}}
 \def\beqa{\begin{eqnarray}}
\def\eeqa{\end{eqnarray}}
\def\eea{\end{array}}
\def\bea{\begin{array}}
\newcommand{\bei}{\begin{itemize}}
\newcommand{\eei}{\end{itemize}}
\newcommand{\bee}{\begin{enumerate}}
\newcommand{\eee}{\end{enumerate}}

\def\1{\openone}

\def\tr{{\rm Tr}}

\def\>{\rangle}
\def\<{\langle}

\def\dt#1{{{\kern -.0mm\rm d}}#1\,}
\def\squareforqed{\hbox{\rlap{$\sqcap$}$\sqcup$}}
\def\qed{\ifmmode\squareforqed\else{\unskip\nobreak\hfil
\penalty50\hskip1em\null\nobreak\hfil\squareforqed
\parfillskip=0pt\finalhyphendemerits=0\endgraf}\fi}

\def\supp{{\rm supp\,}}

\newtheorem{lemma}{Lemma}
\newtheorem{theorem}[lemma]{Theorem}
\newtheorem{main result}[lemma]{Main result}
\newtheorem{proposition}[lemma]{Proposition}
\newtheorem{definition}{Definition}

\newtheorem{fact}[lemma]{Fact}

\def\bep{\begin{proposition}}
\def\eep{\end{proposition}}
\def\bel{\begin{lemma}}
\def\eel{\end{lemma}}

\def\bet{\begin{theorem}}
\def\eet{\end{theorem}}
\def\bed{\begin{definition}}
\def\eed{\end{definition}}
\def\bef{\begin{fact}}
\def\eef{\end{fact}}

\begin{document}

\title{Objectivity in the Photonic Environment Through State Information Broadcasting}

\author{J.~K.~Korbicz}
 \email{jaroslaw.korbicz@ug.edu.pl}
  \affiliation{Institute of Theoretical Physics and Astrophysics, University of Gda\'nsk, 80-952 Gda\'nsk, Poland}
 \affiliation{National Quantum Information Centre in Gdan\'sk, 81-824 Sopot, Poland}
\author{P.~Horodecki}
\affiliation{Faculty of Applied Physics and Mathematics, Gda\'nsk University of Technology, 80-233 Gda\'nsk, Poland}
\affiliation{National Quantum Information Centre in Gdan\'sk, 81-824 Sopot, Poland}
\author{R.~Horodecki}
\affiliation{Institute of Theoretical Physics and Astrophysics, University of Gda\'nsk, 80-952 Gda\'nsk, Poland}
\affiliation{National Quantum Information Centre in Gdan\'sk, 81-824 Sopot, Poland}

\date{\today}

\begin{abstract}
Recently, the emergence of classical objectivity as a property of a quantum state has been explicitly derived for a small 
object embedded in a photonic environment  in terms of a \emph{spectrum broadcast form}---a specific classically correlated state, 
redundantly encoding information about the preferred states of the object in the 
environment. However, the environment was in a pure state and the fundamental problem 
was how generic and robust is the conclusion. 
Here we prove that  despite of the initial environmental noise the emergence of 
the broadcast structure still holds, leading to the perceived objectivity of the state of 
the object. We also show how  this leads  to a quantum Darwinism-type condition,
reflecting classicality of proliferated information in terms of a limit 
behavior of the mutual information. Quite surprisingly, we find ,,singular points'' of the 
decoherence, which can be used to faithfully broadcast a specific classical message through the noisy environment.
\end{abstract}

\keywords{decoherence, quantum darwinism, state broadcasting}

\maketitle

Uninterrupted series of successes of quantum mechanics supports a belief that quantum formalism applies to all of physical reality. 
Thus, in particular, the objective classical world of everyday experience should emerge naturally from the formalism. 
This has been a long-standing problem, already present from the very dawn of quantum mechanics \cite{old_bohr, old_heis}. 
Recently, a crucial step was made in a series of works (see e.g. \cite{ZurekNature,ZurekPRA06,ZwolakZurek}) 
introducing quantum Darwinism---a refined model of decoherence \cite{decoh}, based on a multiple 
environments paradigm: A quantum system of interest $S$ interacts with multiple environments $E_1,\dots,E_N$ instead of just one. 
The authors assumed \cite{ZurekPRA06} that each of these independent fractions effectively measures the system and argued that after the 
decoherence  (with some timescale $\tau_D$) it carries nearly complete classical information about the system, meaning that the information propagates in the environment with a huge redundancy. 
A further step was made in \cite{object} by dropping any explicit  assumptions on the dynamics and applying a operational definition 
of objectivity \cite{ZurekNature} directly to the post-decoherence quantum state. This, together with the Bohr's criterion  of non-disturbance \cite{EPR}, 
allowed to derive a universal state structure---\emph{spectrum broadcast form} (cf. \cite{my}), responsible for the appearance of classical objectivity
in a model- and dynamics-independent way \cite{object}:

\emph{There appears an objectively existing state of the system  $S$ if
the time-asymptotic joint quantum state of $S$ and the observed fraction of the environment $fE$ is of a 
spectrum broadcast form:
\begin{equation}\label{br2}
\varrho_{S:fE}(\infty)=\sum_i p_i \ket{\vec x_i}\bra{\vec x_i}\otimes \varrho^{E_1}_i\otimes\cdots\otimes \varrho^{E_{fN}}_i,\ 
\varrho^{E_k}_i\varrho^{E_k}_{i'\ne i}=0,
\end{equation}
with $\{\ket{\vec x_i}\}$ a pointer basis \cite{ZurekPRD}, $p_i$'s initial pointer probabilities, 
and $\varrho^{E_1}_i,\dots,\varrho^{E_{fN}}_i$ some states of the environments $E_1,\dots,E_{fN}$
with mutually orthogonal supports.}

\begin{figure}[b]
\begin{center}
\includegraphics[scale=0.25]{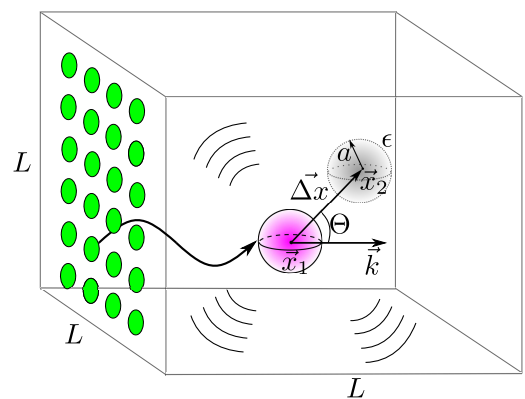}
\caption{The illuminated sphere model \cite{JoosZeh}.
Green dots represent the photons, which constitute the environments $E_1,\dots,E_N$ of the sphere.
The sphere and the photons are enclosed
in a large cubic box of edge $L$; photon momentum eigenstates $\ket{\vec k}$ obey
the periodic boundary conditions.}
\label{general}
\end{center}
\end{figure}

The states (\ref{br2}) "work" by faithfully encoding the same classical information about the system
(index $i$) in each portion of the environment---they describe redundant proliferation (broadcasting) of information,
necessary for objectivity \cite{ZurekNature,object}. A process of formation of  a state (\ref{br2}) is what we call \emph{state information broadcasting} \cite{object}.
It is a weaker form of the quantum state broadcasting \cite{broadcasting,my}.

In this Letter we apply the above novel results to the celebrated model of a dielectric sphere illuminated by photons  \cite{JoosZeh,GallisFleming,HornbergerSipe,ZurekPRL,RiedelZurek}
to show how an objectively existing state of a system \cite{ZurekNature,object} is actually formed  in a course of the quantum evolution with 
a general (not only thermal) noisy environment.
In contrast to the earlier studies
\cite{ZurekPRL,RiedelZurek}, we show it directly
on the fundamental level of quantum states, proving the emergence of the broadcast structure (\ref{br2}),
rather than using information-theoretical conditions, which
so far are only known to necessary, while their sufficiency
is still not known \cite{object}. We thus prove robustness and a generic character of the emergence of 
objectivity---a well known property of the everyday life.
In other words, state information broadcasting process still works even if
the environment is noisy, which in principle might cover or
mismatch proliferation of emerging classical information about the system.  
Moreover,  with a help of the classical Perron-Frobenius Theorem \cite{PF} we show a surprising effect 
of how the decoherence mechanism can be used to \emph{faithfully broadcast} a specific message into the environment. 

\emph{The model \cite{JoosZeh}}.  
A dielectric sphere $S$ of radius $a$ and relative permittivity $\epsilon$ is 
bombarded by a constant flux of photons, constituting the sphere's environment; see Fig.~\ref{general}.
The sphere can be at two possible locations $\vec x_1$ and $\vec x_2$
and the photons are assumed not energetic enough to individually resolve the displacement $\Delta x\equiv|\vec x_2-\vec x_1|$:
\be\label{soft}
k\Delta x\ll 1,
\ee
where $\hbar k$ is some characteristic momentum, but they are able to do so collectively:
If the sphere is initially in a superposition of the localized states $\ket{\vec x_1}$, $\ket{\vec x_2}$,
the scattering photons will localize it via collisional decoherence \cite{JoosZeh}. Here we show that during this process
a broadcast state (\ref{br2}) is formed for the radiation being initially a general mixture of plane waves, concentrated around (\ref{soft}):
\be
\label{mu}
\varrho^{ph}_0=\sum_{\vec k} p(\vec k) \ket{\vec k}\bra{\vec k},\ \supp p\in\{\vec k:|\vec k|\Delta x\ll 1\}
\ee
(in the previous studies only thermal states were considered \cite{JoosZeh,ZurekPRL,RiedelZurek}).
Following \cite{JoosZeh,GallisFleming,ZurekPRL,RiedelZurek},  we use  box normalization
to describe the photons; see Fig.~\ref{general}. We remove it through the thermodynamic limit 
(signified by $\cong$) \cite{ZurekPRL,RiedelZurek}: $V\to\infty$, $N\to\infty$, $N/V=\text{const}$.
The interaction time $t$ enters through the number of scattered photons 
up to time $t$  (a ''macroscopic time'');
see Fig.~\ref{general}:
\be\label{Nt}
N_t\equiv L^2\frac{N}{V}ct,
\ee
where $c$ is the speed of light. We will work with a fixed $t$ and
pass to the decoherence limit $t/\tau_D\to \infty$ (denoted by $\approx$ or $\infty$) at the very end.
The sphere-photons interaction is of a controlled-unitary type (\emph{symmetric environments}):
\be\label{U}
U_{S:E}(t)\equiv\sum_{i=1,2}\ket{\vec x_i}\bra{\vec x_i}\otimes \underbrace{{\bf S}_i\otimes\dots\otimes {\bf S}_i}_{N_t},
\ee
where (assuming  translational invariance) 
${\bf S}_i\equiv {\bf S}_{\vec x_i}=e^{-i\vec x_i\cdot \vec{\hat k}} {\bf S}_0 e^{i\vec x_i\cdot \vec{\hat k}}$ 
is the scattering matrix (see e.g. \cite{Messiah}).

\emph{Macrofractions}. We introduce a crucial \emph{environment coarse-graining}:
we divide the full photonic environment into a number of \emph{macroscopic fractions}, 
each containing $mN_t$ photons, $0\le m \le 1$. By {\it macroscopic} 
we will understand ''scaling with the total number of photons $N_t$''. 
By definition, these are the environment fractions accessible to independent observers, 
searching for an objective state of the sphere \cite{object}. 
In typical situations, detectors used to monitor the environment, 
e.g. eyes, have some minimum detection thresholds and the fractions $mN_t$ 
are meant to reflect it. The concrete fraction size 
is irrelevant here---it is enough that it scales with $N_t$ \cite{macro}.
The detailed initial product state of the environment $(\varrho^{ph}_0)^{\otimes N_t}$
can thus be trivially rewritten as: 
\ben
\underbrace{\varrho^{ph}_0\otimes\dots\otimes\varrho^{ph}_0}_{N_t}&=&\underbrace{\varrho^{ph}_0\otimes\dots\otimes\varrho^{ph}_0}_{mN_t}\otimes\dots\otimes
\underbrace{\varrho^{ph}_0\otimes\dots\otimes\varrho^{ph}_0}_{mN_t}\nonumber\\
&\equiv& \underbrace{\varrho_0^{mac}\otimes\dots\otimes\varrho^{mac}_0}_{M},\label{init_mac}
\een
where $M\equiv 1/m$ is the number of macrofractions and $\varrho_0^{mac}\equiv (\varrho^{ph}_0)^{\otimes mN_t}$ is the initial state of each of them.

\emph{Formation of the broadcast state}. After all the $N_t$ photons have scattered and $(1-f)M$, $0\leq f\leq 1$, macrofractions went unobserved
(the necessary loss of information), the post-scattering
''out''-state  $\varrho_{S:fE}(t)\equiv \tr_{(1-f)E} U_{S:E}(t)\varrho_{S:E}(0)U_{S:E}(t)^\dagger$, is given 
from (\ref{U},\ref{init_mac}) for a product initial state $\varrho_{S:E}(0)\equiv\varrho^S_0\otimes(\varrho^{ph}_0)^{\otimes N_t}$  by:
\ben\label{SfE}
&&\varrho_{S:fE}(t)=\sum_{i=1,2}\langle\vec x_i |\varrho^S_0\, \vec x_i\rangle\ket{\vec x_i}\bra{\vec x_i}\otimes\left[\varrho_i^{mac}(t)\right]^{\otimes fM}\label{i=j}\\
&&+\sum_{i\ne j}\langle\vec x_i |\varrho^S_0\, \vec x_j\rangle\left(\tr {\bf S}_i\varrho^{ph}_0 {\bf S}_j^\dagger\right)^{(1-f)N_t}\ket{\vec x_i}\bra{\vec x_j}\otimes\nonumber\\
&&\qquad\qquad\qquad\qquad\qquad\qquad\otimes\left({\bf S}_i\varrho^{ph}_0 {\bf S}_j^\dagger\right)^{\otimes fN_t}
\label{i ne j},
\een
where $\label{rho_i}\varrho_i^{mac}(t)\equiv\left({\bf S}_i\varrho^{ph}_0 {\bf S}_i^\dagger\right)^{\otimes mN_t}$, $i=1,2$.
We demonstrate that in the soft scattering sector (\ref{soft}), 
the above state approaches asymptotically the broadcast form (\ref{br2}) 
by showing that for $t\gg \tau_D$:
\begin{enumerate}
\item The post-scattering coherent part $\varrho_{S:fE}^{i\ne j}(t)$, defined by (\ref{i ne j}), vanishes in the trace norm (\emph{decoherence}):
\be\label{znikaogon}
||\varrho_{S:fE}^{i\ne j}(t)||_{\text{tr}}\equiv\tr\sqrt{\left[\varrho_{S:fE}^{i\ne j}(t)\right]^\dagger\varrho_{S:fE}^{i\ne j}(t)}\approx 0.
\ee
\item The post-scattering macro-states $\varrho_i^{mac}(t)$ become \emph{perfectly distinguishable}:
$\varrho_1^{mac}(t)\varrho_2^{mac}(t)\approx 0$,
or equivalently using the generalized overlap \cite{Fuchs}:
\ben
&&B\left[\varrho^{mac}_1(t),\varrho^{mac}_2(t)\right]\equiv\nonumber\\
&&\quad\quad\quad\equiv\tr\sqrt{\sqrt{\varrho^{mac}_1(t)}
\varrho^{mac}_2(t)\sqrt{\varrho^{mac}_1(t)}}\approx 0,\label{nonoverlap_norm}
\een
despite of the individual photon states (micro-states) becoming equal in the thermodynamic limit.
\end{enumerate}
The first mechanism above is the usual decoherence of $S$ by $fE$. 
Some form of quantum correlations may still survive it, 
since the resulting state (\ref{i=j}) is generally of a Classical-Quantum (CQ) form \cite{CQ},
but they are damped by the second mechanism and  $\varrho_{S:fE}(\infty)$ becomes a spectrum broadcast state (\ref{br2}) for
$p_i=\langle\vec x_i |\varrho^S_0\, \vec x_i\rangle$.

The decoherence mechanism alone (\ref{znikaogon}) has been extensively studied in the model
for thermal initial states $\varrho^{ph}_0$ (see. e.g. \cite{JoosZeh, GallisFleming, ZurekPRL, RiedelZurek, HornbergerSipe}).
We recall that the decay of the off-diagonal part $\varrho_{S:fE}^{i\ne j}(t)$, defined by (\ref{i ne j}), is governed
by the decoherence factor $|\tr {\bf S}_1\varrho^{ph}_0 {\bf S}_2^\dagger|$, since
$||\varrho_{S:fE}^{i\ne j}(t)||_{\text{tr}}=2|\langle\vec x_1 |\varrho^S_0\, \vec x_2\rangle|
\left|\tr {\bf S}_1\varrho^{ph}_0 {\bf S}_2^\dagger\right|^{(1-f)N_t}.$
For pure $\varrho^{ph}_0$, it reads in the regime (\ref{soft}) \cite{JoosZeh, GallisFleming, ZurekPRL, RiedelZurek, HornbergerSipe}:
\ben\label{psipsi}
&&\langle\vec k|{\bf S}_2^\dagger {\bf S}_1\vec k\rangle=1+i\frac{8\pi\Delta x k^5 \tilde a^6}{3L^2} \cos\Theta\nonumber\\
&&-\frac{2\pi\Delta x^2 k^6 \tilde a^6}{15L^2}
\left(3+11\cos^2\Theta\right)+O\left[\frac{(k\Delta x)^3}{L^2}\right],
\een
where $\Theta$ is the angle between  $\vec k$ and $\vec{\Delta x}\equiv\vec x_2-\vec x_1$,
$\tilde a\equiv a [(\epsilon-1)/(\epsilon+2)]^{1/3}$, 
while for a general distribution (\ref{mu}) it is given in the leading order in $1/L$ by \cite{JoosZeh,ZurekPRL,RiedelZurek, suppl}:
\ben
&&\left|\tr {\bf S}_1\varrho^{ph}_0 {\bf S}_2^\dagger\right|^{(1-f)N_t}\cong\nonumber\\
&&\left[1-\frac{2\pi\Delta x^2 \tilde a^6}{15L^2}\sum_{\vec k}p(\vec k)k^6
\left(3+11\cos^2\Theta_{\vec k}\right)\right]^{(1-f)N_t}\label{bareta}\\
&&\xrightarrow{\text{therm.}}\text{exp}\left[-\frac{(1-f)}{\overline{\tau_D}}t\right],\label{decay_ogon_mix}
\een
where $\overline{\tau_D}^{\;-1}\equiv\frac{2\pi}{15}\frac{N}{V}\Delta x^2 c \tilde a^6 
\int d\vec k p(\vec k) k^6  \left(3+11\cos^2\Theta_{\vec k}\right)$
is the decoherence time.

Completing the step (\ref{nonoverlap_norm}) is more involved.  For the 
micro-states $\varrho^{mic}_i\equiv {\bf S}_i\varrho_0^{ph}{\bf S}_i^\dagger$ we obtain under (\ref{soft}) \cite{suppl}:
\ben
B\left(\varrho^{mic}_1,\varrho^{mic}_2\right) = 1-\frac{\bar\eta-\bar\eta'}{L^2}\xrightarrow{\text{therm.}}1,
\label{ort_micmix}
\een
where:
\ben
&&\bar\eta\equiv \frac{L^2}{2}\left(1-\sum_{\vec k}p(\vec k)\left|\langle\vec k|{\bf S}_1^\dagger {\bf S}_2\vec k\rangle\right|^2\right)
\cong \left(\overline{\tau_D}\frac{N}{V}c\right)^{-1}\label{eta}\\
&&\bar\eta'\equiv \frac{L^2}{2}\sum_{\vec k}\sum_{\vec k'\ne \vec k}p(\vec k)\left|\langle\vec k|{\bf S}_1^\dagger {\bf S}_2\vec k'\rangle\right|^2,
\label{eta'}
\een
so that $\varrho^{mic}_{1,2}$ become equal 
and encode no information about $S$. Same holds if the observed portion $\mu$ of the environment $E$ is {\it microscopic}, i.e. not scaling with $N_t$:
\ben
&&\varrho_{S:\mu E}(0)=\varrho^S_0\otimes\left(\varrho_0^{mac}\right)^{\otimes \mu} \xrightarrow[\text{therm.}]{t\gg\tau_D}
\varrho_{S:\mu E}(\infty)=\nonumber\\
&& \left(\sum_{i=1,2}p_i\ket{\vec x_i}\bra{\vec x_i}\right)
\otimes\left[\varrho^{mic}\right]^{\otimes \mu}\label{product}.
\een
This is a ''product phase'' \cite{object}, in which the mutual information $I[\varrho_{S:\mu E}(\infty)]=0$.

Passing to macro-states 
the situation changes as now:
\ben
&&B\left[\varrho^{mac}_1(t),\varrho^{mac}_2(t)\right]=
\left(\tr\sqrt{\sqrt{\varrho^{mic}_1} \varrho^{mic}_2\sqrt{\varrho^{mic}_1}}\right)^{mN_t}\nonumber\\
&&\cong \left(1-\frac{\alpha\bar\eta}{L^2}\right)^{mN_t}\xrightarrow{\text{therm.}}
\text{exp}\left[-\frac{\alpha m}{\overline{\tau_D}}t\right],\label{ort_mix}
\een
where $\alpha\equiv(\bar\eta-\bar\eta')/\bar\eta$ \cite{RiedelZurek}
and  (\ref{eta}) was used.
Thus, whenever $\alpha\ne 0$ ($\alpha=0$ e.g. for an isotropic illumination \cite{suppl}),
$B[\varrho^{mac}_1(t),\varrho^{mac}_2(t)]\approx 0$ for $t\gg\overline{\tau_D}/\alpha$, despite (\ref{ort_micmix}), 
i.e. the macro-states become perfectly distinguishable via orthogonal projectors on their supports.
The latter are contained in  $\text{span}\{\ket{\vec k} :\vec k\in \supp p\}^{\otimes mN_t}$ (cf. (\ref{mu})),
rotated by ${\bf S}_1^{\otimes mN_t}$ and ${\bf S}_2^{\otimes mN_t}$ respectively.  
Eqs.~(\ref{decay_ogon_mix},\ref{ort_mix}) together imply an asymptotic formation of the spectrum broadcast state (\ref{br2}):
\ben
&&\varrho_{S:fE}(0)=\varrho^S_0\otimes\left[\varrho_0^{mac}\right]^{\otimes fM} 
\xrightarrow[\text{therm.}]{t\gg\tau_D}\varrho_{S:fE}(\infty)=\nonumber\\
&&\sum_{i=1,2}\langle\vec x_i |\varrho^S_0\, \vec x_i\rangle\ket{\vec x_i}\bra{\vec x_i}\otimes\left[\varrho_i^{mac}(\infty)\right]^{\otimes fM}
\label{b-state}
\een
with  $\varrho^{mac}_1(\infty)\varrho^{mac}_2(\infty)=0$ \cite{corr}. The scattering (\ref{b-state}) is thus a combination of the localization measurement in the pointer basis
$\ket{\vec x_i}$ and spectrum broadcasting of the result, described by a CC-type channel \cite{my} :
\ben
\Lambda^{S\to fE}_\infty(\varrho_0^S)\equiv\sum_i\langle\vec x_i |\varrho^S_0\, \vec x_i\rangle\left[\varrho_i^{mac}(\infty)\right]^{\otimes fM}.
\label{L}
\een 
As a consequence of (\ref{b-state}), it follows that \cite{suppl}:
\be\label{QD}
I[\varrho_{S:fE}(t)] \xrightarrow[\text{therm.}]{t\gg\tau_D} H_S\ \text{for any}\ 0<f<1,
\ee
i.e. the mutual information 
becomes asymptotically independent of the fraction size $f$ (as long as it is macroscopic). 
This is the entropic objectivity condition of quantum Darwinism, leading to the characteristic classical plateau \cite{ZurekNature}.
We stress that here (\ref{QD}) is \emph{derived} as a consequence of the state information broadcasting (\ref{b-state}) and we call
this regime a "broadcasting phase" \cite{object}. When the whole $E$ is observed, modulo a micro-fraction,
there appears from (\ref{i ne j}) a "full information phase", when quantum correlations are retained and $I[\varrho_{S:fE}(t)]\approx I_{max}$.

Comparing (\ref{decay_ogon_mix}) and (\ref{ort_mix}) we observe that, unlike in the pure case \cite{object},  the timescales of decoherence (\ref{znikaogon})
and distinguishability (\ref{nonoverlap_norm}) are a priori different (cf. \cite{RiedelZurek}): $\overline{\tau_D}$ and $\overline{\tau_D}/\alpha$ respectively.
Since $0\leq\alpha\leq 1$, the broadcast state is fully formed for $t \gg \overline{\tau_D}/\alpha$.
Environment noise thus slows down the  formation of the broadcast state \cite{Holevo}.

\emph{"Singular points" of decoherence}. 
Let the initial state of the sphere have a diagonal representation
$\varrho_0^S=\sum_i\lambda_{0i}\ket{\phi_i}\bra{\phi_i}$. Then, in (\ref{b-state})
there appears a  stochastic matrix $P_{ij}(\phi)\equiv|\langle\phi_i|\vec x_j\rangle|^2$, which
by the Perron-Frobenius Theorem \cite{PF} possesses at least one stable probability distribution $\lambda_{*i}(\phi)$:
$\sum_j P_{ij}(\phi)\lambda_{*j}(\phi)=\lambda_{*i}(\phi)$. It exists
for \emph{any} initial eigenbasis $\ket{\phi_i}$. Let us choose it as the initial spectrum:
$\lambda_{0i}\equiv\lambda_{*i}(\phi)$. Then, the scattering (\ref{b-state}) not only leaves this distribution unchanged, but 
broadcasts it into the environment:
\ben
&&\left[\sum_i\lambda_{*i}(\phi)\ket{\phi_i}\bra{\phi_i}\right]\otimes\left(\varrho_0^{mac}\right)^{\otimes fM} 
\xrightarrow[\text{therm.}]{t\gg\tau_D}\varrho_{S:fE}(\infty)=\nonumber\\
&&=\sum_{i}\left(\sum_jP_{ij}(\phi)\lambda_{*j}(\phi)\right)\ket{\vec x_i}\bra{\vec x_i}
\otimes\left(\varrho_i^{mac}\right)^{\otimes fM}\nonumber\\
&&=\sum_{i}\lambda_{*i}(\phi)\ket{\vec x_i}\bra{\vec x_i}
\otimes\left(\varrho_i^{mac}\right)^{\otimes fM}.
\een
The initial spectrum does not ''decohere''. 
This surprising Perron-Frobenius broadcasting \cite{my}, 
can thus be  used to faithfully (in the asymptotic limit above) broadcast the classical message $\{\lambda_{*i}(\phi)\}$ through the
environment macro-fractions, however noisy they are.

\emph{Final remarks.} There is one straightforward generalization to 
many parties. Consider several spheres, each with its own 
photonic environment, separated by distances $D$, 
$kD\gg1$ (cf. (\ref{soft})). The interaction is then a product of (\ref{U}), e.g. for two spheres
$U_{S_1S_2:E_1E_2}(t)\equiv\sum_{i,j=1,2}\ket{\vec x_i}\bra{\vec x_i}\otimes\ket{\vec y_j}\bra{\vec y_j}
\otimes{\bf S}_i^{\otimes N_t}\otimes\tilde{\bf S}_j^{\otimes N_t}$, 
where $\vec x_i,\vec y_j$ are the spheres' positions and ${\bf S}_i,\tilde{\bf S}_j$ are the corresponding scattering matrices.
The asymptotic state (\ref{b-state}) provides objectivisation of classical correlations \cite{my}, e.g. 
$p_{ij}\equiv\langle \vec x_i,\vec y_j|\varrho_0^S\vec x_i,\vec y_j\rangle$, 
measurable by observers who have an access to photons originating from all the spheres.

In the studied model, as in the majority of decoherence models, 
the system-environment interaction is of a  form:
\be\label{Hint}
H_{int}=g A_S \sum_{k=1}^N X_{E_k},
\ee
where $g$ is a coupling constant and $A_S,X_{E_1},\dots,X_{E_N}$ are some observables on the system and the environments respectively. 
The eigenbasis of $A=\sum_i a_i \ket i\bra i$ becomes the pointer basis---it is arguably put by hand by the choice
of the form (\ref{Hint}).  
It is then an interesting question if there are more general interactions, without an a priori priviledged basis,
which nevertheless lead to an asymptotic formation of spectrum broadcast states (\ref{br2}).

Finally, it would be extremely interesting to test our findings
experimentally. In fact,  our central object, the broadcast
state (\ref{br2}), is in principle directly observable 
through e.g. quantum state tomography---a well developed, successful, and widely used technique \cite{tomo}.

\acknowledgements
This work is supported by the ERC Advanced Grant QOLAPS and  
National Science Centre project Maestro DEC-2011/02/A/ST2/00305. We thank W. H. Zurek and C. J Riedel for discussions. 
P.H. and R.H. acknowledge discussions with K. Horodecki, M. Horodecki, and K. \.Zyczkowski.

\appendix
\section{Supplemental Material}
\subsection{The decoherence factor for mixed states}
Here we calculate  $\left|\tr {\bf S}_1\varrho^{ph}_0 {\bf S}_2^\dagger\right|$ in the leading order in the box size $1/L$
for the mixed states (\ref{mu}):
\ben
&&\left|\tr {\bf S}_1\varrho^{ph}_0 {\bf S}_2^\dagger\right|^2=
\sum_{\vec k,\vec k'} p(\vec k)p(\vec k')\langle \vec k|{\bf S}_2^\dagger{\bf S}_1\vec k\rangle\langle \vec k'|{\bf S}_1^\dagger{\bf S}_2\vec k'\rangle\nonumber\\
&&\cong\sum_{\vec k,\vec k'} p(\vec k)p(\vec k')\left(1+\frac{iA_{\vec k}}{L^2}-\frac{B_{\vec k}}{L^2}\right)
\left(1-\frac{iA_{\vec k'}}{L^2}-\frac{B_{\vec k'}}{L^2}\right)\nonumber\\
&&\cong\sum_{\vec k,\vec k'} p(\vec k)p(\vec k')\left(1+\frac{iA_{\vec k}}{L^2}-\frac{iA_{\vec k'}}{L^2}-\frac{B_{\vec k}}{L^2}-\frac{B_{\vec k'}}{L^2}\right)\\
&&=1-2\sum_{\vec k}p(\vec k)\frac{B_{\vec k}}{L^2}, 
\een
where we used Eq.~(\ref{psipsi}) keeping only the terms up to $1/L^2$ and introduced
$A_{\vec k}\equiv\frac{8}{3}\pi\Delta x k^5 \tilde a^6\cos\Theta_{\vec k}$, 
$B_{\vec k}\equiv \frac{2\pi}{15}\Delta x^2 k^6 \tilde a^6
\left(3+11\cos^2\Theta_{\vec k}\right)$. This gives in the leading order:
\ben
\left|\tr {\bf S}_1\varrho^{ph}_0 {\bf S}_2^\dagger\right|
=\sqrt{1-2\sum_{\vec k}p(\vec k)\frac{B_{\vec k}}{L^2}}\cong 1-\sum_{\vec k}p(\vec k)\frac{B_{\vec k}}{L^2},\nonumber\\
\een
leading to Eq.~(\ref{bareta}).

\subsection{Calculation of $B\left(\varrho^{mic}_1,\varrho^{mic}_2\right)$}
We calculate the Bhattacharyya coefficient \cite{Fuchs_suppl} $B\left(\varrho^{mic}_1,\varrho^{mic}_2\right)$ (defined in Eq.~(\ref{nonoverlap_norm})) for the individual photon 
states (micro-states) $\varrho^{mic}_i\equiv {\bf S}_i\varrho_0^{ph}{\bf S}_i^\dagger$
for general momentum-diagonal initial states (\ref{mu}) (our calculation is partially similar to that of Ref.~\cite{RiedelZurek}). Let:
\be
\sqrt{\varrho^{mic}_1} \varrho^{mic}_2\sqrt{\varrho^{mic}_1}\equiv {\bf S}_1\left(\sum_{\vec k, \vec k''} M_{\vec k \vec k''} 
\ket{\vec k}\bra{\vec k''}\right){\bf S}_1^\dagger,\label{M0}
\ee
where we have introduced a matrix:
\ben
M_{\vec k \vec k''} \equiv \sqrt{p(\vec k)p(\vec k'')}\sum_{\vec k'}p(\vec k') \langle \vec k| {\bf S}_1^\dagger {\bf S}_2 \vec k'\rangle
\langle \vec k'| {\bf S}_2^\dagger {\bf S}_1 \vec k''\rangle\nonumber.\\
\label{M}
\een
By Eq.~(\ref{mu}) it is supported in the sector (\ref{soft}), and we diagonalize it in the leading order in $1/L$. 
For that, we first decompose matrix elements $M_{\vec k \vec k''}$ in $1/L$ and keep the leading terms only. 
Let us write:
\be\label{b}
{\bf S}_1^\dagger {\bf S}_2={\bf 1}-({\bf 1}-{\bf S}_1^\dagger {\bf S}_2)\equiv {\bf 1}-b.
\ee
Matrix elements of $b$ between vectors satisfying  (\ref{soft}) are of the order of $1/L$ at most: i)
the diagonal elements  $b_{\vec k\vec k}=1-\langle\vec k|{\bf S}_1^\dagger {\bf S}_2\vec k\rangle=O(1/L^2)$ by Eq.~(\ref{psipsi}); 
ii) the off-diagonal elements are determined by the unitarity of ${\bf S}_1^\dagger {\bf S}_2$ and the order of the diagonal ones:
$1=|\langle\vec k| {\bf S}_1^\dagger {\bf S}_2 \vec k\rangle|^2+\sum_{\vec k'\ne \vec k}|\langle\vec k |{\bf S}_1^\dagger {\bf S}_2 \vec k'\rangle|^2
=1-O(1/L^2)+\sum_{\vec k'\ne \vec k}|b_{\vec k\vec k'}|^2$  for any fixed $\vec k$ satisfying (\ref{soft})
(there is a single sum here), where we again used Eq.~(\ref{psipsi}). Hence:
\be\label{offdiag}
\forall_{\vec k} \colon \sum_{\vec k'\ne \vec k}|b_{\vec k\vec k'}|^2=\sum_{\vec k'\ne \vec k}\left|\langle\vec k|{\bf S}_1^\dagger {\bf S}_2\vec k'\rangle\right|^2
=O\left(\frac{1}{L^2}\right). 
\ee
As a byproduct, from (\ref{offdiag}) it follows also that in the energy sector (\ref{soft}), ${\bf S}_1\cong {\bf S}_2$ in the strong operator topology:
$||({\bf S}_1-{\bf S}_2)|\phi\rangle||^2=||b\ket{\phi}||^2\xrightarrow{\text{therm.}}0$
for any $\ket\phi$ from the subspace defined by (\ref{soft}). 
From Eqs.~(\ref{psipsi},\ref{M},\ref{b},\ref{offdiag}) we obtain in the leading order:
\ben
M_{\vec k \vec k''}&=&p(\vec k)^2\delta_{\vec k\vec k''}-p(\vec k)^{3/2}\sqrt{p(\vec k'')}b^*_{\vec k''\vec k}\nonumber\\
&-&p(\vec k'')^{3/2}\sqrt{p(\vec k)}b_{\vec k\vec k''}+ O\left(\frac{1}{L^4}\right).\label{M'}
\een
The first term is non-negative and is of the order of unity, while the rest is of the order $1/L$ and forms a Hermitian matrix.
We can thus calculate the desired eigenvalues $m(\vec k)$ of $M_{\vec k \vec k''}$ using 
standard, stationary perturbation theory of quantum mechanics,
treating the terms with the matrix $b$ as a small perturbation.
Assuming a generic non-degenerate situation (the measure $p(\vec k)$ in Eq.~(\ref{mu}) is injective), we obtain:
\ben\label{m}
m(\vec k)=p(\vec k)^2\left(1-b^*_{\vec k\vec k}-b_{\vec k\vec k}\right)+O\left(\frac{1}{L^4}\right),
\een
and finally:
\ben
&&B\left(\varrho^{mic}_1,\varrho^{mic}_2\right)=\tr\sqrt{\sqrt{\varrho^{mic}_1} \varrho^{mic}_2\sqrt{\varrho^{mic}_1}}
=\tr\sqrt M\nonumber\\
&& \cong\sum_{\vec k}p(\vec k)\sqrt{1-2\text{Re}b_{\vec k\vec k}}\cong\sum_{\vec k}p(\vec k)\left(1-\text{Re}b_{\vec k\vec k}\right)\\
&&=1+\frac{1}{2}\sum_{\vec k}\left(\frac{M_{\vec k\vec k}}{p(\vec k)}-p(\vec k)\right)=\frac{1}{2}
+\sum_{\vec k}\frac{p(\vec k)}{2}\left|\langle\vec k|{\bf S}_1^\dagger {\bf S}_2\vec k\rangle\right|^2\nonumber\\
&&+\sum_{\vec k}\sum_{\vec k'\ne \vec k}\frac{p(\vec k)}{2}\left|\langle\vec k|{\bf S}_1^\dagger {\bf S}_2\vec k'\rangle\right|^2
\equiv 1-\frac{\bar\eta-\bar\eta'}{L^2},\label{rhorho}
\een
where we have used Eqs.~(\ref{M0},\ref{m},\ref{psipsi},\ref{M}) in the respective order, and introduced:
\ben
&&\bar\eta\equiv \frac{L^2}{2}\left(1-\sum_{\vec k}p(\vec k)\left|\langle\vec k|{\bf S}_1^\dagger {\bf S}_2\vec k\rangle\right|^2\right)
\cong \left(\overline{\tau_D}\frac{N}{V}c\right)^{-1}\label{eta_suppl}\\
&&\bar\eta'\equiv \frac{L^2}{2}\sum_{\vec k}\sum_{\vec k'\ne \vec k}p(\vec k)\left|\langle\vec k|{\bf S}_1^\dagger {\bf S}_2\vec k'\rangle\right|^2
\label{eta'_suppl}
\een
(in Eq.~(\ref{eta_suppl}) we have used Eqs.~(\ref{bareta},\ref{decay_ogon_mix})).
We note that $\bar\eta,\bar\eta'$ are of the order of unity in $1/L$ by Eqs.~(\ref{decay_ogon_mix},\ref{offdiag}).

\subsection{Isotropic illumination}
We show that for a completely isotropic illumination the macro-states never orthogonalize and hence 
such an environment carries no information on the sphere's localization, as it intuitively should not (cf. Refs.~\cite{ZurekPRL_suppl,ZwolakZurek_suppl}). 
To prove it, we calculate 
\be
\sum_{\vec k,\vec k'}p(\vec k)\left|\langle\vec k|{\bf S}_1^\dagger {\bf S}_2\vec k'\rangle\right|^2
\ee
for $p(\vec k)\equiv p(k)(1/4\pi)$.
Or more precisely, since we are working in the box normalization, the measure  
is 
\be
p(\vec k)\equiv p(k)(1/\Omega_k), 
\ee
where $\Omega_k$ is the number of the discrete box states $\ket{\vec k}$ 
with the fixed length $k=|\vec k|$. In the continuous limit $\Omega_k$ approaches $4\pi k^2$.
As the scattering is by assumption elastic, matrix elements $\langle\vec k|{\bf S}_1^\dagger {\bf S}_2\vec k'\rangle$ are non-zero
only for the equal lengths $k=k'$ and hence:
\be\label{sectors}
{\bf S}_1^\dagger {\bf S}_2\equiv\bigoplus_k U_k,\  U_k^\dagger U_k={\bf 1}_k.
\ee
Decomposing the summations over  $\vec k,\vec k'$ into the sums over the lengths $k,k'$ and the directions $\vec n(k), \vec n(k')$
and using (\ref{sectors}), we obtain:
\ben
&&\sum_{\vec k,\vec k'}p(\vec k)\left|\langle\vec k|{\bf S}_1^\dagger {\bf S}_2\vec k'\rangle\right|^2=\nonumber\\
&&\sum_k \frac{p(k)}{\Omega_k}\sum_{\vec n(k),\vec n(k')}\left|\langle\vec k|{\bf S}_1^\dagger {\bf S}_2\vec k'\rangle\right|^2=\\
&&\sum_k \frac{p(k)}{\Omega_k}\tr\left(P_k {\bf S}_1^\dagger {\bf S}_2 P_k {\bf S}_2{\bf S}_1^\dagger\right)=\\
&&\sum_k p(k)\frac{\tr P_k}{\Omega_k}=1,\label{1}
\een 
where $P_k\equiv\sum_{\vec n(k)}\ket{\vec k}\bra{\vec k}$ is a projector onto the subs-space of a fixed length $k$, and hence $\tr P_k=\Omega_k$. 
Comparing with Eq.~(\ref{rhorho}),  Eq.~(\ref{1}) leads to $\bar\eta-\bar\eta'=0$, and since by definition $\alpha=(\bar\eta-\bar\eta')/\bar\eta$, to $\alpha=0$. 

\subsection{Derivation of the quantum Darwinism  relation (\ref{QD})}
\label{IHS}
We show that Eq.~(\ref{QD}) follows from the
mechanisms of i) decoherence, Eq.~(\ref{znikaogon}), and ii) distinguishability, Eq.~(\ref{nonoverlap_norm})
and is thus a consequence of the state information broadcasting process.

We generalize to mixed environment states the similar calculations from Ref.~\cite{object-s}. 
Let the post-interaction $S:fE$ state for a fixed, finite box $L$ and time $t$ be $\varrho_{S:fE}(L,t)$.
It is given by Eqs.~(\ref{i=j},\ref{i ne j}) and now we 
explicitly indicate the dependence on $L$ in the notation. Then:
\ben
&&\left|H_S-I\left[\varrho_{S:fE}(L,t)\right]\right|\leq\nonumber\\
&&\left|I\left[\varrho_{S:fE}(L,t)\right]-I\left[\varrho^{i=j}_{S:fE}(L,t)\right]\right|\label{coh}\\
&&+\left|H_S-I\left[\varrho^{i=j}_{S:fE}(L,t)\right]\right|,\label{ort}
\een
where $\varrho^{i=j}_{S:fE}(L,t)$ is the decohered part of $\varrho_{S:fE}(L,t)$, given by Eq.~(\ref{i=j}).
We first bound the difference (\ref{coh}), decomposing the mutual information using conditional information 
$S_{\text{vN}}(\varrho_{S:fE}|\varrho_{fE})\equiv S_{\text{vN}}(\varrho_{S:fE})-S_{\text{vN}}(\varrho_{fE})$:
\be
I(\varrho_{S:fE})=S_{\text{vN}}\left(\varrho_S\right)-S_{\text{vN}}\left(\varrho_{S:fE}|\varrho_{fE}\right),
\ee
so that:
\ben
&&\left|I\left[\varrho_{S:fE}(L,t)\right]-I\left[\varrho^{i=j}_{S:fE}(L,t)\right]\right|\leq\nonumber\\
&&\left|S_{\text{vN}}\left[\varrho_{S}(L,t)\right]-S_{\text{vN}}\left[\varrho^{i=j}_{S}(L,t)\right]\right|+\label{SS}\\ 
&&\Big|S_{\text{vN}}\left[\varrho_{S:fE}(L,t)\big|\varrho_{fE}(L,t)\right]\nonumber\\
&&\quad\quad\quad\quad\quad -S_{\text{vN}}\left[\varrho_{S:fE}^{i=j}(L,t)\Big|\varrho_{fE}^{i=j}(L,t)\right]\Big|.
\label{Scond}
\een
From Eq.~(\ref{soft}), the 
total $S:fE$ Hilbert space is finite-dimensional for a finite $L,t$: there are $fN_t=$$fL^2(N/V)ct$ photons (cf. Eq.~(\ref{Nt}))
and the number of modes of each photon is approximately $(4\pi/3)(L/2\pi\Delta x)^3$. 
Hence, the total dimension is $2\times L^2f(N/V)ct\times(1/6\pi^2)(L/\Delta x)^3<\infty$ and we can use the
Fannes-Audenaert \cite{FAd} and the Alicki-Fannes \cite{FannesAlicki} inequalities to bound
(\ref{SS}) and (\ref{Scond}) respectively. For (\ref{SS}) we obtain:
\ben
&&\left|S_{\text{vN}}\left[\varrho_{S}(L,t)\right]-S_{\text{vN}}\left[\varrho^{i=j}_{S}(L,t)\right]\right|\nonumber\\
&&\leq\frac{1}{2}\epsilon_E(L,t)\log(d_S-1)+h\left[\frac{\epsilon_E(L,t)}{2}\right],
\een
where $h(\epsilon)\equiv-\epsilon\log\epsilon-(1-\epsilon)\log(1-\epsilon)$ is the binary Shannon entropy
and:
\ben
&&\epsilon_E(L,t)\equiv ||\varrho_{S}(L,t)-\varrho^{i=j}_{S}(L,t)||_{tr}\\
&&=||\varrho^{i\ne j}_{S}(L,t)||_{tr}\cong 2|c_{12}|\left[1-\frac{1}{c\overline{\tau_D}L^2}\left(\frac{N}{V}\right)^{-1}\right]^{L^2\frac{N}{V}ct}
\label{ELt}
\een
with $c_{12}\equiv\langle\vec x_1 |\varrho^S_0\, \vec x_2\rangle$, 
where we have used the same reasoning (\ref{decay_ogon_mix}),
but with $f=0$. For (\ref{Scond}) the same reasoning and the Alicki-Fannes inequality give:
\ben
&&\left|S_{\text{vN}}\left[\varrho_{S:fE}(L,t)\big|\varrho_{fE}(L,t)\right]
-S_{\text{vN}}\left[\varrho_{S:fE}^{i=j}(L,t)\big|\varrho_{fE}^{i=j}(L,t)\right]\right|\nonumber\\
&&\leq4\epsilon_{fE}(L,t)\log d_S+2 h\left[\epsilon_{fE}(L,t)\right],
\een
with:
\ben
\epsilon_{fE}(L,t)&\equiv& ||\varrho_{S:fE}(L,t)-\varrho^{i=j}_{S:fE}(L,t)||_{tr}\\
&=&||\varrho^{i\ne j}_{S:fE}(L,t)||_{tr}\\
&\cong& 2|c_{12}|\left[1-\frac{1}{c\overline{\tau_D}L^2}\left(\frac{N}{V}\right)^{-1}\right]^{L^2(1-f)\frac{N}{V}ct}.\label{EfLt}
\een
Above $L,t$ are big enough so that $\epsilon_E(L,t), \epsilon_{fE}(L,t)<1$. 
Eqs.~(\ref{SS}-\ref{EfLt}) give an upper bound on the difference (\ref{coh}) in terms of the 
decoherence speed (\ref{znikaogon}).

To bound the "orthogonalization" part (\ref{ort}) (see Ref.~\cite{ZwolakZurek_suppl} for a related analysis), 
we note that since $\varrho^{i=j}_{S:fE}(L,t)$ is a CQ-state
(cf. Eq.~(\ref{i=j})), its mutual information is given by the Holevo quantity \cite{Holevo_suppl}:
\ben
I\left[\varrho^{i=j}_{S:fE}(L,t)\right]=\chi\left\{p_i,\varrho_i^{mac}(t)^{\otimes fM}\right\}.
\een
From the Holevo Theorem it is bounded by \cite{Holevo}:
\be\label{hol}
I_{max}(t)\leq\chi\left\{p_i,\varrho_i^{mac}(t)^{\otimes fM}\right\}\leq H\left(\{p_i\}\right)\equiv H_S,
\ee
where $I_{max}(t)\equiv \max_{\mathcal E}I[p_i\pi^{\mathcal E}_{j|i}(t)]$ is the fixed time maximal mutual information, extractable 
through generalized measurements $\{\mathcal E_j\}$ on the  ensemble 
$\{p_i,\varrho_i^{mac}(t)^{\otimes fM}\}$, and the conditional probabilities read:
\be\label{piE}
\pi^{\mathcal E}_{j|i}(t)\equiv\tr[\mathcal E_j\varrho_i^{mac}(t)^{\otimes fM}]
\ee
(here and below $i$ labels the states, while $j$ the measurement outcomes).
We now relate $I_{max}(t)$ to the generalized overlap   
$B\left[\varrho_1^{mac}(t)^{\otimes fM},\varrho_2^{mac}(t)^{\otimes fM}\right]$ (cf. Eq.~(\ref{nonoverlap_norm})),
which we have calculated in Eq.~(\ref{ort_mix}). Using the method of Ref.~\cite{Fuchs_suppl},
slightly modified to unequal a priori probabilities $p_i$, we obtain for an arbitarry measurement $\mathcal E$:
\ben
&&I\left(\pi^{\mathcal E}_{j|i}p_i\right)=I\left(\pi^{\mathcal E}_{i|j}\pi^{\mathcal E}_j\right)
=H\left(\{p_i\}\right)-\sum_{j=1,2}\pi^{\mathcal E}_jh\left(\pi^{\mathcal E}_{1|j}\right)\nonumber\\
&&\\
&&\geq H\left(\{p_i\}\right)-2\sum_{j=1,2}\pi^{\mathcal E}_j \sqrt{\pi^{\mathcal E}_{1|j}\left(1-\pi^{\mathcal E}_{1|j}\right)}\\
&&=H\left(\{p_i\}\right)-2\sqrt{p_1p_2}\sum_{j=1,2}\sqrt{\pi^{\mathcal E}_{j|1}\pi^{\mathcal E}_{j|2}},
\een
where we have first used Bayes Theorem $\pi^{\mathcal E}_{i|j}=(p_i/\pi^{\mathcal E}_j)\pi^{\mathcal E}_{j|i}$,
$\pi^{\mathcal E}_j\equiv\sum_i\pi^{\mathcal E}_{j|i}p_i=\tr(\mathcal E_j\sum_i\varrho_i)$, then the fact that we have only two states:
$\pi^{\mathcal E}_{2|j}=1-\pi^{\mathcal E}_{1|j}$, so that $H(\pi^{\mathcal E}_{\cdot|j})=h(\pi^{\mathcal E}_{1|j})$,
and finally $h(p)\leq2\sqrt{p(1-p)}$. On the other hand, 
$B(\varrho_1,\varrho_2)=\min_{\mathcal E}\sum_j\sqrt{\pi^{\mathcal E}_{j|1}\pi^{\mathcal E}_{j|2}}$ \cite{Fuchs_suppl}. Denoting the optimal
measurement by $\mathcal E_*^B(t)$ and recognizing that $H(\{p_i\})=H_S$, we obtain:
\ben
&&I_{max}(t)\geq I\left[p_i\pi^{\mathcal E_*^B(t)}_{j|i}(t)\right]\geq H_S-\\
&&-2\sqrt{p_1p_2}\,B\left[\varrho_1^{mac}(t)^{\otimes fM},\varrho_2^{mac}(t)^{\otimes fM}\right]\\
&&=H_S-2\sqrt{p_1p_2}\,B\left[\varrho^{mac}_1(t),\varrho^{mac}_2(t)\right]^{fM} 
\een
Inserting the above into the bounds (\ref{hol}) gives the desired upper bound on the difference (\ref{ort}):
\ben
&&\left|H_S-I\left[\varrho^{i=j}_{S:fE}(L,t)\right]\right|\leq
2\sqrt{p_1p_2}\,B\left[\varrho^{mac}_1(t),\varrho^{mac}_2(t)\right]^{fM}\nonumber\\
&&
\een
where the generalized overlap is given by Eq.~(\ref{ort_mix}): 
\ben
&&B\left[\varrho^{mac}_1(t),\varrho^{mac}_2(t)\right]\cong\nonumber\\
&&\quad\quad\quad\left[1-\frac{\alpha}{c\overline{\tau_D}L^2}\left(\frac{N}{V}\right)^{-1}\right]^{L^2m\frac{N}{V}ct}.
\label{Bh}
\een

Gathering all the above facts together finally leads to a bound on   
$\left|H_S-I\left[\varrho_{S:fE}(L,t)\right]\right|$ in terms of the speed of 
i) decoherence (\ref{znikaogon}) and ii) distinguishability (\ref{nonoverlap_norm}):
\ben
&&\left|H_S-I\left[\varrho_{S:fE}(L,t)\right]\right|\leq h\left[\frac{\epsilon_E(L,t)}{2}\right]+
 2h\left[\epsilon_{fE}(L,t)\right]+\nonumber\\
&&\label{gen1}\\
&&4\epsilon_{fE}(L,t)\log 2+2\sqrt{p_1p_2}\,B\left[\varrho^{mac}_1(t),\varrho^{mac}_2(t)\right]^{fM},\label{gen2}
\een
where $\epsilon_E(L,t)$, $\epsilon_{fE}(L,t)$, $B\left[\varrho^{mac}_1(t),\varrho^{mac}_2(t)\right]$
are given by Eqs.~(\ref{ELt}), (\ref{EfLt}), and (\ref{Bh}) respectively. Choosing
$L,t$ big enough so that $\epsilon_E(L,t),\epsilon_{fE}(L,t)\leq 1/2$ (when the binary 
entropy $h(\cdot)$ is monotonically increasing), we remove the unphysical box and obtain an estimate on
the speed of convergence of $I\left[\varrho_{S:fE}(L,t)\right]$ to $H_S$:
\ben
&&\lim_{L\to\infty}\left|H_S-I\left[\varrho_{S:fE}(L,t)\right]\right|\leq 
h\left(|c_{12}|\text e^{-\frac{t}{\overline{\tau_D}}}\right)\\
&&+2h\left(2|c_{12}|\text e^{-\frac{(1-f)}{\overline{\tau_D}}t}\right)
+8|c_{12}|\text e^{-\frac{(1-f)}{\overline{\tau_D}}t}\log 2\\
&&+2\sqrt{p_1p_2}\text e^{-\frac{\alpha f}{\overline{\tau_D}}t}.
\een
This finishes the derivation of the Quantum Darwinism condition (\ref{QD}).

We note that the result (\ref{gen1},\ref{gen2}) is in fact a general statement, valid 
in any model where: i) the system $S$ is effectively a qubit; ii) the system-environment
interaction is of a environment-symmetric, controlled-unitary type:
\begin{theorem}
Let a two-dimensional quantum system $S$ interact with $N$ identical environments, each described by
a finite-dimensional Hilbert space, through a controlled-unitary interaction:
\be
U(t)\equiv\sum_{i=1,2}\ket i\bra i\otimes U_i(t)^{\otimes N}.
\ee
Let the initial state be $\varrho_{S:E}(0)=\varrho_0^S\otimes(\varrho_0^E)^{\otimes N}$ 
and $\varrho_{S:E}(t)\equiv U(t)\varrho_{S:E}(0)U(t)^\dagger$. Then for any $0<f<1$ and $t$ big enough:
\ben
&&\left|H(\{p_i\})-I\left[\varrho_{S:fE}(t)\right]\right|\leq h\left[\frac{\epsilon_E(t)}{2}\right]+
 2h\left[\epsilon_{fE}(t)\right]+\nonumber\\
\\
&&4\epsilon_{fE}(t)\log 2+2\sqrt{p_1p_2}\,B\left[\varrho^{mac}_1(t),\varrho^{mac}_2(t)\right]^{fN},\label{gen3}
\een
where: 
\ben
&&p_i\equiv\langle i|\varrho_0^S|i\rangle,\,\varrho_i(t)\equiv U_i(t)\varrho_0^EU_i(t)^\dagger,\\
&&\epsilon_E(t)\equiv ||\varrho_{S}(t)-\varrho^{i=j}_{S}||_{tr},\\
&&\epsilon_{fE}(t)\equiv ||\varrho_{S:fE}(t)-\varrho^{i=j}_{S:fE}(t)||_{tr}.
\een
\end{theorem}


\begin{thebibliography}{99}
\bibitem{corr} By (\ref{decay_ogon_mix},\ref{ort_mix}), the asymptotic formation of the
spectrum broadcast state (\ref{b-state}) relies on the full product form of the initial state $\varrho_{S:E}(0)$ and the interaction (\ref{U}) in each block $i$.
However, one can allow for correlated/entangled fractions of photons, as long as they stay microscopic, i.e. do not scale with $N_t$.
The corresponding terms then factor out in front of the exponentials in (\ref{decay_ogon_mix},\ref{ort_mix})
and the formation of the spectrum broadcast states is not affected.
\end{thebibliography}
\end{document}